\documentstyle[prb,aps,epsfig]{revtex}

\begin{document}
\title
{
{\bf Spin diffusion and relaxation in three-dimensional
isotropic Heisenberg antiferromagnets}\\
}
\author
{
K.A.Kikoin,
M.N.Kiselev
}
\address
{
RRC "Kurchatov Institute", 123182, Moscow, Russia 
}
\date{\today}
\maketitle

\begin{abstract}
A theory is proposed for kinetic effects in isotropic Heisenberg
antiferromagnets at temperatures above the Neel point. A metod
based on the analysis of a set of Feynman diagrams for the  kinetic 
coefficients is developed for studying the critical dynamics.
The scaling behavior of the generalized coefficient of
spin diffusion and relaxation constant in the paramagnetic phase
is studied in terms of the approximation of coupling modes. It is shown
that the kinetic coefficients in an antiferromagnetic system are singular
in the fluctuation region. The corresponding critical indices for
diffusion and relaxation processes are calculated. The scaling 
dimensionality of the kinetic coefficients agrees with the predictions
of dynamic scaling theory and a renormalization group analysis.
The proposed theory can be used to study the momentum and frequency 
dependence of the kinetic parameters, and to determine the form of 
the scaling functions. The role of nonlocal correlations and spin-liquid
effects in magnetic systems is briefly discussed.
\\
\mbox{}
\\
\end{abstract}

\section{Introduction}
\par
In recent years there had been considerable activity, both
experimental and theoretical, in the field of critical phenomena of
antiferromagnets with Heisenberg and RKKY interactions \cite{Chan89} -\cite{Mil93}.
This increase of interest \cite{Mil93}-\cite{HF2} was stimulated by new experiments 
in high $T_C$ superconductors and heavy fermion compounds.
In particular, the
critical spin fluctuations are supposed to be responsible for Non-Fermi-Liquid
low temperature behavior of the heat capacity and resistivity in
$CeCu_{6-x}Au_x$ and $Ce_{1-x}La_xRu_2Si_2$ \cite{HF1},\cite{HF2}
in the vicinity of the quantum critical point.
Besides, the  Fermi-type 
Resonating Valence Bond excitations were introduced by \cite{And1},\cite{And2}
in the framework of the Heisenberg model  to describe unusual magnetic 
properties of high $T_c$ cuprates \cite{And2}
and Ce - based heavy fermion compounds \cite{KKP92},\cite{KKM94}.
In this case the critical spin fluctuations were shown to be 
important in the spin liquid formation mechanism.
However, the scaling behavior of kinetic coefficients in the
presence of spin - liquid fluctuations can be differ
significantly from the predictions of  the dynamic scaling theory
\cite{KKM97}.

In this paper we present a microscopic description of the scaling
behavior of spin diffusion and relaxation coefficients in
isotropic Heisenberg antiferromagnet in the critical region above 
the Neel temperature. The scaling dimension of kinetic coefficients
was predicted by Halperin and Hohenberg \cite{Halp1}-\cite{Halp2}.
They proposed the
dynamic scaling hypothesis based on the assumption that the
critical exponents of kinetic coefficients are the same on both
sides of the phase transition point.  
The microscopic treatment of the spin diffusion in the paramagnetic
phase of ferromagnet was proposed by S.V.Maleev
\cite{Mal1} - \cite{Mal2}. Within the framework of this approach
the approximations which are
necessary for validity of the dynamic scaling hypothesis  were established
and momentum and frequency dependence of the spin diffusion coefficient
was analyzed. We will follow Maleev's approach to derive the scaling 
behavior of kinetic coefficients for Heisenberg antiferromagnet.

The qualitative picture of the critical phenomena expressed
in terms of wave number $k$ and inverse coherence 
length $\xi^{-1}$ is well known \cite{Halp1}-\cite{Halp2}.
There are two regions: "hydrodynamic" regime deals with the longitudinal
fluctuations $q\xi \ll 1$ of the order parameter - the staggered
magnetization ${\bf N}={\bf N}_1-{\bf N}_2$ and "critical" one
considers the fluctuations with  $q\xi \gg 1$
(the momentum  ${\bf q}=|{\bf k-Q}|$ 
corresponds to small deviation of the momentum from the AF vector). 
However, there is an additional conserved quantity in AFM which is the 
total magnetization ${\bf M}={\bf M}_1+{\bf M}_2$. The fluctuations of 
this parameter do not lead to any singularities in spin correlators.
Nevertheless, there is a region of longitudinal fluctuations
of $M$ $k\xi \ll 1$ which  is also can be called as "hydrodynamic".
The goal of our work is to investigate the spin correlation functions 
behavior in the paramagnetic phase of AF and to establish the relations 
between the kinetic coefficients in two fluctuation regions of the
phase diagram.

The dynamic of magnetization 
fluctuations is described by the van Hove microscopic diffusion equation
in the long wave limit
\begin{equation}
\frac{\partial{\bf M}}{\partial t} = D_0 \nabla^2 {\bf M},
\label{0.1}
\end{equation}
where  $D_0$ is the spin diffusion coefficient. 
This equation is derived from the conservation low for the
total magnetic moment,
since the  operator ${\bf M}$ commutes with the
Hamiltonian. The fluctuation behavior of nonconserved order parameter is
radically different. According to the general postulates of the 
thermodynamics of weakly nonequilibrium processes \cite{PaPok82}
the rate of change ${\bf N}$
is proportional to the conjugate thermodynamical force 
\begin{equation}
\frac{\partial{\bf N}}{\partial t} = -\frac{\Gamma_0}{G_0} {\bf N},
\label{0.2}
\end{equation}
here $G_0$ - is the static susceptibility,
and the kinetic coefficient $\Gamma_0$ satisfies the condition 
$\Gamma_0 >0$.
The relaxation function determined by the equation
(\ref{0.1}) is nearly uniform, therefore the gradient terms for relaxation
processes are omitted. In contrast to relaxation
the first nonzero contribution to the spin diffusion equation is
proportional to  $q^2$. Although the average value
of ${\bf M}$ is equal to zero on the both sides of the
phase transition, the fluctuations of total
magnetization around zero take place. However, 
unlike the FM case the diffusion mode is not the "critical"
mode for AF transition.

Thus, we are interested in the dynamical susceptibility of a cubic
Heisenberg antiferromagnet in zero magnetic field above the Neel temperature,
\begin{equation}
H= - \sum_{<i,j>} V_{ij} {\bf S}_i{\bf S}_j
\label{0.0}
\end{equation}
and the dipolar interaction \cite{Mal2} is neglected.
As is  known the susceptibility is related to  the retarded Green's
function by the following equality:
\begin{equation}
\chi({\bf k}, \omega) = (g\mu_0)^2 K^R_{SS}({\bf k},\omega),
\label{0.3}
\end{equation}
where $g$ is Lande splitting factor, $\mu_0$ is Bohr magneton and
\begin{equation}
K^R_{SS}({\bf k},\omega) = i \int_0^{\infty} dt e^{i\omega t}
<[S^z_{\bf k}(t),S^z_{\bf -k}(0)]>,
\label{0.4}
\end{equation}

$$
{\bf S}_{\bf k}=\frac{1}{\sqrt{N}}\sum_i e^{-i{\bf k}{\bf R}_i} {\bf S}_i ,
$$

$$
{\bf M} = <{\bf S}_0 >, \;\;\;\;\;\;{\bf N} = <{\bf S}_{\bf Q_{AFM}} >.
$$
Using the equations (\ref{0.1}, \ref{0.2}) one can obtain an expression for the
retarded Green functions in the fluctuation region in the following form:
\begin{equation}
K^R_{SS}({\bf k \to 0},\omega)=
{\cal K}({\bf k},\omega)= G_0(k)\frac{iDk^2}{\omega + iDk^2}
\label{0.5}
\end{equation}
\begin{equation}
K^R_{SS}({\bf q}=({\bf k}- {\bf Q}) \to 0,\omega)
={\cal L}({\bf q},\omega)=
\frac{1}{-i\omega/\Gamma + G_0^{-1}(q)}
\label{0.6}
\end{equation}
Here $G_0$ is the static spin correlation function. The equations (\ref{0.5},
\ref{0.6}) correspond to diffusion and relaxation regimes respectively.

When $\tau =|T-T_c|/T_c \ll {\sf Gi}$ (${\sf Gi}$ is the Ginzburg number which characterizes
the conditions when the Landau theory is valid) the fluctuations become
important. The dynamic properties in this region can be described by Halperin -
Hohenberg scaling hypothesis. According to this theory the equation
for the dynamic susceptibility can be expressed in terms of the scaling 
function $F$:
\begin{equation}
K^R_{SS}({\bf k},\omega) = 
G_0({\bf k}) F( k\xi,\frac{\omega}{T_c\tau^{\nu z}}),
\label{0.7}
\end{equation}
Thus, the dynamic critical exponent $z$, which characterizes the critical 
fluctuations energy scale
$\omega \sim k^z$, is connected with the static critical exponent
$\nu \approx 2/3$ defined by the expression $\xi \sim \tau^{-\nu}$. 
We assume the static behavior of susceptibility in the form
$G_0(q) \sim \xi^{2-\eta}$.
We  also neglect the Fisher index $\eta$ which determines the "anomalous
dimension" \cite{PaPok82}. This approximation is valid in 3D case
\cite{PaPok82}.
To derive the scaling properties of AF 
one have to introduce two types of the scaling functions:
${\cal F}_1$ and ${\cal F}_2$ 
\begin{equation}
{\cal K}({\bf k},\omega) = 
G_0({\bf k}) {\cal F}_1( k\xi,\frac{\omega}{T_c\tau^{\nu 
z}}),\;\;\;\;\;\;
{\cal L}({\bf q},\omega) = 
G_0({\bf q}) {\cal F}_2( q\xi,\frac{\omega}{T_c\tau^{\nu z}}),
\label{0.8}
\end{equation}
In principal, expressions (\ref{0.7}, \ref{0.8}) 
determine the kinetic coefficients
$D_0$ and $\Gamma_0$ in terms of coherence length. Moreover,
the renormalization group analyses \cite{Halp1},\cite{FreMa75}
has shown that the kinetic
coefficients are singular in the fluctuation region of AF.

The theory under consideration is a variant of the mode-mode
coupling theory proposed by Kawasaki \cite{Kaw76} (see also
\cite{Lov1}-\cite{Lov3}).
We extend to AF systems the dynamic scaling method offered by Maleev
for ferromagnets.   As was mentioned above, the aim of our paper
is to derive the scaling behavior and momentum and frequency
dependence of ${\cal F}$ functions in the fluctuation region of AF.
Assumptions which are crucial for the microscopic substantiations
of dynamic scaling hypothesis will be also formulated.

\section{Generalized kinetic coefficients}
We consider the dynamic susceptibility of  cubic
Heisenberg antiferromagnet in the fluctuation region above the Neel
temperature. The equations (\ref{0.5}, \ref{0.6}) 
can be rewritten in more general form:
\begin{equation}
K^R_{SS}({\bf k },\omega)= \frac{i\gamma({\bf k}, \omega)}
{\omega + iG_0^{-1}({\bf k})\gamma({\bf k}, \omega)}
\label{1.1}
\end{equation}
There is a simple expression for the spin - diffusion coefficient
in the hydrodynamical regime  
\begin{equation}
D_0=\lim_{{\bf k}\to 0} \lim_{\omega \to 0}
k^{-2}\gamma({\bf k}, \omega)G_0^{-1}({\bf k}),
\label{1.2}
\end{equation}
The generalized kinetic coefficient in the relaxation region is
$\gamma({\bf k}, \omega)=\Gamma({\bf k}, \omega)$. The different 
limits $k \to 0, \omega \to 0$ in the expressions
(\ref{0.5}, \ref{0.6}) 
significantly depend on the relations between $k$ and $\omega$.
These limits are completely analogous to those considered in the 
Fermi Liquid theory \cite{AGD62}.
In the treatment below we consider the  the quasistatic limit of
${\bf k} \to 0$, $|\omega|/{\bf k}^2 \to 0$.

As it was shown by Maleev, it is possible to determine the kinetic
coefficients in terms of the Kubo functions for
the operators $S$ and $\dot{S}$, where $\dot{S} =dS/dt$,  
\begin{equation}
\gamma({\bf k}, \omega)=\frac{\Phi_{\dot{S}\dot{S}}({\bf k}, \omega)}
{1+G_0^{-1}(k)\Phi_{\dot{S}S}({\bf k}, \omega)},
\label{1.3}
\end{equation}
and
$$
\Phi_{AB}({\bf k}, \omega)=\frac{1}{i\omega}
[K^R_{AB}({\bf k}, \omega)-K^R_{AB}({\bf k},0 )],
$$
$$
K^R_{AB}({\bf k}, \omega)=i \int_0^{\infty} dt e^{i\omega t}
<[A_{\bf k}(t),B_{\bf -k}(0)]>.
$$
The formal expression (\ref{1.3}) is exact, and it takes into
account the nonlinear character of the relaxation forces.
In the case of purely exchange interaction, $\dot{S} \sim k$ and
$\gamma = \Phi_{\dot{S}\dot{S}}({\bf k}, \omega)$. The Kubo
function in the denominator vanishes at ${\bf k=0}$, and we arrive
at the expression $\gamma = \Phi_{\dot{S}\dot{S}}$.
However, in studying the dispersion of the kinetic coefficients,
i.e. its dependence on $k$, $\omega$, we cannot, generally speaking,
neglect this function in the denominator.

Using the properties of spin operators, it is not difficult to obtain the 
relations between different retarded spin Green's functions
$K^R_{SS}({\bf k}, 
\omega)$, $K^R_{\dot{S}S}({\bf k}, \omega)$ and $K^R_{\dot{S}\dot{S}}({\bf k}, 
\omega)$ 
in the paramagnetic phase. These relations are result of dispersion 
relations \cite{Izs88}:
$$
K^R_{\dot{S}S}({\bf k}, \omega)=-i\omega K^R_{SS}({\bf k}, 
\omega),\;\;\;\;
K^R_{S\dot{S}}({\bf k}, \omega)=-K^R_{\dot{S}S}({\bf k}, \omega)= 
i\omega K^R_{SS}({\bf k}, \omega),\;\;\;\;
$$
\begin{equation}
\omega^2K^R_{SS}({\bf k}, \omega)=
[K^R_{\dot{S}\dot{S}}({\bf k}, \omega)-K^R_{\dot{S}\dot{S}}({\bf k},0 )]
\label{1.4}
\end{equation}
In particular, it  is seen from equation (\ref{1.4}) that
the function  
$K^R_{\dot{S}\dot{S}}({\bf k}, \omega)$ possesses the same symmetry
properties as spin - spin correlation function.

The spin current operators taken in the form
\begin{equation}
K_{\dot{S}\dot{S}}({\bf k}, \omega_n)=\frac{(a^2T_c\alpha)^2}{6N}
\int_0^{1/T} d\tau e^{i\omega_n \tau}
\sum_{\bf p_1,p_2}(\nabla V({\bf p_1}) {\bf k})(\nabla V({\bf p_2}) {\bf 
k})
<T_\tau (S^\mu_{\bf p_1 +k}S^\rho_{-\bf p_1})_\tau
(S^\mu_{\bf -p_2 -k}S^\rho_{\bf p_2})_0>
\label{1.6}
\end{equation}
can be connected with the Kubo functions for Matsubara frequencies
$\omega_n$ with the use of equation of motion:

\begin{equation}
\dot{S}^\alpha_{\bf k}=
-\frac{1}{\sqrt{N}}\sum_{\bf p}
[V({\bf p}+{\bf k})-V({\bf p})]
\epsilon_{\alpha\beta\gamma}S^\beta_{{\bf p}+{\bf k}}
S^\gamma_{-{\bf p}}
\label{1.5}
\end{equation}

The spin - current correlators can be calculated
diagrammatically and then the explicit equations
(\ref{1.1}-\ref{1.3}) for the kinetic coefficients can be obtained.
When deriving eq. (\ref{1.6}), we have confined ourselves to
the lowest terms of the expansion in powers of $ka$ and have
put $\nabla V({\bf p}) \approx {\bf p} T_c a^2\alpha$, where
$\alpha \sim 1$. Since, we are interested in
the correction terms of order $(k\xi)^2$ and $(k\xi)^3$; we
consistently neglect the $(ka)^2$ and $(ka)^3$ corrections,
because $\xi \gg a$.
Thus the problem of derivation of the kinetic coefficients
is reduced to the analysis of four - spin correlators with current vertices.
The later problem can be
solved by analytical continuation of the temperature
Feynman's diagram to the upper half-plane of complex variable $\omega$
The Feynman diagrams corresponding to the 
spin-current correlation function are shown on figure 1. 

The "bare" poles of the Green's functions (\ref{0.5}, \ref{0.6}) 
lie on imaginary axis.
Introducing fictitious quasi - particles "diffusons" and "relaxons"
we obtain the self-consistent equations for kinetic coefficients and  determine
their scaling dimensions.

The static susceptibility in the fluctuation region
obeys the  Ornstein - Zernike law
\begin{equation}
G_0({\bf q})=K^R_{SS}({\bf q},0)=\frac{A}{T_c\tau^{2\nu}}
\frac{1}{(q\xi)^2+1},
\label{1.7}
\end{equation}
(here A is a constant ($A \sim 1$), $\tau \ll 1$). 
There is no singularities of static susceptibility in the diffusion region
$k\xi \ll 1$, where $G_0\approx A/(2T_c)$. 

In Sec. III the relations between kinetic coefficients are established
and scaling dimensions of kinetic coefficient are analyzed.
Sec. IV is devoted to a momentum and frequency dependence of spin diffusion
and spin relaxation constants.

\section{Relations between kinetic coefficients}
We begin with analyzing the  set of diagrams which cannot be separated
into two parts by cutting one line of interaction, and   
include these diagrams into the irreducible self-energy part of the
spin - current correlation function. Using the definition of
$\gamma$ and properties of the functions $K$ we
obtain an expression for generalized kinetic coefficient
in terms of irreducible self - energy parts:
\begin{equation}
\gamma({\bf k},\omega)=\frac{1}{i\omega}
[\Sigma^R_{\dot{S}\dot{S}}({\bf k},\omega)-
\Sigma^R_{\dot{S}\dot{S}}({\bf k},0)
+\frac{{\cal R}^R_{\dot{S}S}({\bf k},\omega)\gamma({\bf k},\omega)
{\cal R}^R_{S\dot{S}}({\bf k},\omega)}{-i\omega + G_0^{-1}(k)\gamma({\bf 
k},\omega)}]
[1+G_0^{-1}\frac{{\cal R}^R_{\dot{S}S}({\bf k},\omega)\gamma({\bf 
k},\omega)}
{i\omega(-i\omega + G_0^{-1}(k)\gamma({\bf k},\omega))}]^{-1}
\label{2.1}
\end{equation}
The equation (\ref{2.1}) can be derived both from analysis of 
diagrammatic series for the 
 spin - current correlators
\cite{Mal1}, and directly from Larkin equation
\cite{KKM94},\cite{Izs88}. Below we  use the following notations
for retarded spin Green's functions: $K^R_{\dot{S}S}({\bf k},\omega)={\cal R}^R({\bf k},\omega) 
K^R_{SS}({\bf k},\omega)$, and $\Sigma^R_{AB}$ - are the irreducible 
self - energies. To obtain the graphical representation for
the irreducible part $\Sigma^R_{\dot{S}\dot{S}}$ one have to replace the
"dressed" vertex in fig.1  by the irreducible one.
The estimation of ${\cal R}$ in the framework of mean - field theory
\cite{Mal1}, \cite{VLP1} results in:
\begin{equation}
{\cal R} \sim (k\xi)(ka) \ll (k\xi)^2
\label{2.1a}
\end{equation}

Moreover, due to analytical properties of spin correlators,
${\cal R}^R \sim \omega$.
We assume that the expression for ${\cal R}$ contains the small parameter
$a/\xi$ in a limit of small $\omega$ not only in critical region and
neglect  this contribution below. Therefore, the generalized
kinetic coefficient $\gamma$ can be rewritten in terms of irreducible self - 
energies only:
\begin{equation}
\gamma({\bf k},\omega)=\frac{1}{i\omega}
(\Sigma^R_{\dot{S}\dot{S}}({\bf k},\omega)-
\Sigma^R_{\dot{S}\dot{S}}({\bf k},0))
\label{2.2}
\end{equation}
The general set of diagrams 
for the irreducible self - energy part
$\Sigma_{\dot{S}\dot{S}}$  in imaginary time
can be classified by the number of intermediate states. First, we
confine ourselves to the  diagrams with two - particles intermediate states
(fig.2)
$$
\Sigma_{\dot{S}\dot{S}}^{(2)}({\bf k},i\omega)=
\frac{(T_c a^2 \alpha)^2}{\sqrt{N}} T\sum_\epsilon \sum_{\bf p}
({\bf k}{\bf \Lambda^{(2)}}({\bf p, k},i\omega,i\epsilon,i(\omega-
\epsilon)))
({\bf k}{\bf \Lambda^{(2)}}^\dagger
({\bf p, k},i\epsilon,i(\omega-\epsilon),i\omega))
\times
$$
\begin{equation}
\times
K_{SS}({\bf p},i\epsilon)K_{SS}({\bf k-p},i\omega-i\epsilon)
\label{2.3}
\end{equation}
Replacing the sum over momentum ${\bf p}$ by the integration one should use
the "cutoff" $p \sim \xi^{-1}$ as an upper limit of integration.
This means that the integrations are restricted by the region close to the
appropriate singularities (small ${\bf p}$ and small ${\bf q}$
for ${\bf p} \sim {\bf q}+{\bf Q}$
in the vicinity of AF vector ${\bf Q}$).
From the point of view of the energy dependence, the vertices
${\bf \Lambda}^{(2)}$ being the  three - point functions are the analytical
functions of all three frequencies $\omega$ and, in each of the frequencies they have 
a cut along the real axis; they have no singularities in the physical
region of variables $\omega$ \cite{Mal3}. Due to these properties it is possible
to separate the vertex into a static part and dynamic contribution,
disappearing in the limit $\omega \to 0$ \cite{Mal3}. 
We concentrate below on the static vertex.

As it can be seen from the fig.2(a), the static vertices correspond to the 
longitudinal processes of "diffuson" - "diffuson" or "relaxon"- relaxon"
pair creation and annihilation, 
since in this case only the modes of the same origin couple together.
On the other hand, the static Green function $G({\bf k})$
is not sensitive to the direction of the momentum. In the other words,
the corresponding lines in the diagrams are not 
directed.  For this reason, the processes of
diffuson and relaxon scattering contain the same vertices as the processes
of pair creation and annihilation. However, for such a scattering processes
in the limit $k \to 0$ the Ward identity holds
\cite{PaPok82},\cite{AGD62} (fig.3):
\begin{equation}
{\bf \Lambda}^{(2)}({\bf p,k},0) \sim \partial G_0^{-1}/ \partial {\bf 
p} 
\label{2.4}
\end{equation}
As was mentioned, the region of integration in the second term is concentrated
near the points ${\bf p} \approx {\bf Q}$. 
The contribution from the staggered magnetization fluctuations into the
spin diffusion can be calculated by means of the  substitution
${\bf p}={\bf q}+{\bf Q}$ and,
by use of the  property $\partial G_0^{-1}/ \partial {\bf p} =\partial G_0^{-1}/ \partial {\bf 
q}$.

Let us consider the diagram fig.2(b). The external momentum 
of this diagram can be taken equal to ${\bf Q}_{AF}$. Therefore, we have two
different coupling modes, "diffuson" and 
"relaxon". Thus, this diagram corresponds to the process of  
"diffuson" - "relaxon" pair creation. For this reason we can not use the Ward
identity to analyze this contribution. However, the "bare" vertex (fig.3)
has the scaling dimension
$$
{\bf \Lambda}^{(2)}_0({\bf p,Q},0) 
\sim \partial V/ \partial {\bf p}  \sim {\bf p}
$$
Besides, in the magnetic Brillouin Zone 
the points 0 and $Q$ are equivalent.
Taking into account the direction independence of interacting modes
in fluctuating region one can assume that the processes of multiple
scattering on the static fields do not influence the scaling dimension
of the static vertex. 
\footnote{As was mentioned above, the index of anomalous dimension 
(Fisher critical exponent) is taken to be zero.}

Using the properties of vertices it is not difficult to perform the summation
over $\epsilon$ and the analytical continuation in $\omega$
\cite{AGD62}. As a result we  obtain the following expressions for kinetic
coefficients
\begin{equation}
D_0^{(2)}=\tilde{A} T_c \int_{-\infty}^{\infty}\frac{d\varepsilon}{2\pi}
coth(\frac{\varepsilon}{2T})\sum_{\bf p}(\nabla G_0^{-1}({\bf p}))^2
[{\it Im}{\cal K}({\bf p},\varepsilon)\frac{\partial}{\partial 
\varepsilon}
{\it Im}{\cal K}({\bf p-k},\varepsilon)+{\it Im}{\cal L}({\bf 
p},\varepsilon)
\frac{\partial}{\partial \varepsilon}{\it Im}{\cal L}({\bf p-
k},\varepsilon)]
\label{2.5}
\end{equation}

\begin{equation}
\Gamma_0^{(2)}=\tilde{B}\int_{-\infty}^{\infty}\frac{d\varepsilon}{2\pi}
coth(\frac{\varepsilon}{2T})\sum_{\bf p}
(\nabla G_0^{-1}({\bf p}){\bf Q})^2\times
[{\it Im}{\cal K}({\bf p},\varepsilon)\frac{\partial}{\partial 
\varepsilon}
{\it Im}{\cal L}({\bf p-q},\varepsilon)+{\it Im}{\cal L}({\bf 
p},\varepsilon)
\frac{\partial}{\partial \varepsilon}{\it Im}{\cal K}({\bf p-
q},\varepsilon)]
\label{2.6}
\end{equation}
There index $^{(2)}$ stands for the processes with only two - particle
intermediate states taken into account. The equations generalize 
the corresponding equation
(\ref{2.6}) are obtained by Maleev \cite{Mal1} 
for ferromagnets. In that case it is possible to retain only
the first term in expression (\ref{2.5}), since one 
deals with a single mode regime.
Equations (\ref{2.5},\ref{2.6}) can be rewritten in a different form.
Taking ${\bf k}=0$ and ${\bf q}=0$  and 
integrating by parts over $\epsilon$ one can obtain the following result:
\begin{equation}
D_0^{(2)}=\frac{\tilde{A}}{4} \int_{-
\infty}^{\infty}\frac{d\varepsilon}{2\pi}
sh^{-2}(\frac{\varepsilon}{2T})\sum_{\bf p}(\nabla G_0^{-1}({\bf p}))^2
[({\it Im}{\cal K}({\bf p},\varepsilon))^2
+({\it Im}{\cal L}({\bf p},\varepsilon))^2]
\label{2.7}
\end{equation}

\begin{equation}
\Gamma_0^{(2)}=\frac{\tilde{B}}{2T_c}
\int_{-\infty}^{\infty}\frac{d\varepsilon}{2\pi}
sh^{-2}(\frac{\varepsilon}{2T})\sum_{\bf p}
(\nabla G_0^{-1}({\bf p}){\bf Q})^2
{\it Im}{\cal K}({\bf p},\varepsilon)
{\it Im}{\cal L}({\bf p},\varepsilon)
\label{2.8}
\end{equation}
 
These expressions are in fact a generalization of equations derived 
by Maleev \cite{Mal1} for the case of two coupling modes.

As it was mentioned above, the main contribution  to the integrals 
(\ref{2.5},\ref{2.6}) appears from the vicinity of scaling 
functions singularities (\ref{0.8}). Besides, the characteristic energies of critical fluctuations
satisfy the condition  $\omega^* \ll T_c$ due to the "critical slowing down".
Thus, we can replace the hyperbolic functions (\ref{2.5} -\ref{2.7})
by their arguments. Calculating integrals
over the frequencies and extracting the scaling dimension of kinetic
coefficients we obtain the relation between spin diffusion coefficient
and the relaxation constant.
\begin{equation}
D_0= b_1 T_c^2 a^4 (\frac{\xi}{a})^{-3}\frac{1}{D_0}+
b_2 T_c a^2 (\frac{\xi}{a})\frac{1}{\Gamma_0}
\label{2.9}
\end{equation}
It is interesting to note that the expressions (\ref{2.9}) can be derived by 
direct 
substitution of retarded Green functions
(\ref{0.5},\ref{0.6}) to the expressions (\ref{2.5},\ref{2.7}).
After integration over the frequencies we obtain the equations on
kinetic coefficients which contain only integration over momentum
of the different powers of the static correlator $G_0$. As a result the first
term is determined by the two - diffuson intermediate state and the second
term corresponds to two - relaxon intermediate state. 

The integrals (\ref{2.6} - \ref{2.8}) can be calculated in the same way.
The equation which determines the relaxation constant in terms of spin
diffusion coefficient can be written in following form:
\begin{equation}
\Gamma_0= c_1 (\frac{\xi}{a}) \frac{1}{\Gamma_0}+
c_2 (\frac{\xi}{a})\frac{D_0/(T_c a^2)}{\Gamma_0^2}
\label{2.10}
\end{equation}

The coefficients $b_{1,2},c_{1,2} \sim 1$ in eqs.
(\ref{2.9},\ref{2.10}) depend on the shape of a scaling
function and, strictly speaking, can not be calculated in the framework
of method used. Solution of self - consistent system of {\it
algebraic} equations leads to following 
scaling dimension of kinetic coefficient
\footnote{
Spin diffusion coefficient
in ferromagnet is not a singular function. It behaves as
$D_0/(T_c a^2) \sim (\xi/a)^{-1/2}$}
\begin{equation}
D_0/(T_c a^2) \sim \Gamma_0 \sim (\xi/a)^{1/2}
\label{2.11}
\end{equation}
Behavior of kinetic coefficient agrees with predictions of dynamic scaling
theory \cite{Halp1}-\cite{Halp2}  and with renormalization group analysis
\cite{Halp2}\cite{FreMa75}. Therefore,  the kinetic coefficients
of AF are singular in fluctuating region, and spin diffusion is determined
by the intermediate relaxation processes. A correction
in  (\ref{2.7}) corresponding to the influence of self - diffusion 
is proportional to
$\delta D_0/D_0 \sim (\xi/a)^{-4} \sim \tau^{8/3} \ll 1$. 
Hence, these processes can be omitted in the critical region. 
Calculation of dynamic
critical exponent z (see eq.\ref{0.7}) results in $z=3/2$.

The simple physical assumptions explaining the
spin diffusion and relaxation in the fluctuation region are based on the 
following argument: in the vicinity of the phase transition point
$T \to T_c$ the regions of short range order are formed. 
The characteristic length scale of the ordered regions is 
$\xi$. The excitations
in these regions are the AF magnons with sound like dispersion law. Estimations
of the spin diffusion coefficient results in   $D_0 \sim \xi^2/t_{diff}$, 
where $t_{diff} \sim \xi / c$ is a characteristic diffusion time and 
$c \sim \xi^{-1/2}$ is a "sound" velocity \cite{Halp1}, 
therefore $D_0 \sim \xi^{1/2}$.
Due to the dynamic scaling hypothesis which postulates the
invariance of the dynamic scaling exponent $z$, the characteristic scaling
behavior of relaxation constant is 
$\Gamma_0 \sim \xi^{1/2}$.

Although the kinetic coefficients are singular, the characteristic
relaxation times of staggered magnetization and spin diffusion
become infinite. This result guarantees
the existence of macroscopic ordered regions, corresponding to the
partial equilibrium  of the system \cite{L10}.

It should be emphasized that we did not use any assumptions about the 
elementary excitations
spectrum in ordered phase when deriving the expressions (\ref{2.9}).
Our study is  based only on a concept of total magnetization 
conservation and  order parameter non-conservation.

\section{Corrections to the kinetic coefficients}

Here we study the kinetic coefficient as a functions
of finite but small ${\bf k}$ and $\omega$. For this sake
we use the relations between spin correlation functions (\ref{1.3})
and Kubo functions (\ref{2.1}). It can be seen that the corrections due to
momentum and frequency dependence of kinetic coefficient are determined
both by the nonlinear character of the relaxation forces
and the momentum and frequency dependence of irreducible self - energy
parts. According to the estimation (\ref{2.1a}) one can derive the
frequency and momentum dependence of spin diffusion and relaxation constants
only in terms of linear response theory. Therefore, we can neglect the
nonlinear character of relaxation forces.

First, we consider the static renormalizations of kinetic coefficients.
It is not difficult to obtain the usual  expansions
for  spin diffusion and relaxation constants in series of
$(k\xi)^{2n}$, $(q\xi)^{2n}$:
$$
D^{(2)}({\bf k},0)=D_0(0,0)[1+\alpha^{'}(k\xi)^2 +...]
$$

$$
\Gamma^{(2)}({\bf q},0)=\Gamma_0(0,0)[1+\beta^{'}(q\xi)^2 +...]
$$
As is known  \cite{Mig68}, these expansion can be done due 
to the singularities of static spin correlation functions
in points ${\bf k}_i= -n^2\xi^{-2}$, where $n$ are integer. The coefficients 
$\alpha^{'}$, $\beta^{'}$ are determined by the shape of static correlation
function.

We analyze the energy dependence of kinetic coefficients.
Using equation (\ref{2.2}) we obtain  
the following expressions for the real and imaginary
parts of the generalized coefficient $\gamma({\bf k},\omega)$:
\begin{equation}
{\it Re} \gamma({\bf k},\omega) = \frac{{\it Im}
\Sigma^R_{\dot{S}\dot{S}}({\bf k},\omega)}{\omega},\;\;\;
{\it Im} \gamma({\bf k},\omega) = 
- \frac{{\it Re}\Sigma^R_{\dot{S}\dot{S}}({\bf k},\omega)-
{\it Re}\Sigma^R_{\dot{S}\dot{S}}({\bf k},0)}{\omega}
\label{3.1}
\end{equation}
Since ${\it Im}\gamma$ is the odd function of $\omega$ and 
${\it Re}\gamma$ is the even function of $\omega$ the first nonzero term
in the energy expansion of kinetic coefficients is proportional to
$\omega^2$.

Let us introduce the effective generalized kinetic coefficient
$\gamma^*$ by the definition:
\begin{equation}
\gamma^* = \frac{\displaystyle\frac{\partial}{\partial \omega} {\it Im}
\Sigma^R_{\dot{S}\dot{S}}({\bf k},\omega)|_{\omega=0}}
{1 + \displaystyle G_0^{-1}({\bf k})\frac{\partial}{\partial (\omega)^2}
{\it Re}\Sigma^R_{\dot{S}\dot{S}}({\bf k},\omega)|_{\omega=0}}
\label{3.2}
\end{equation}

The expression for the coefficient $\gamma^*$ is completely analogous to the
definition of  effective mass in the theory of quantum liquids. The 
renormalization constant $Z$ is similar to the  Fermi - liquid $Z$ factor.
$$
Z=\frac{1}
{1 + \displaystyle G_0^{-1}({\bf k})\frac{\partial}{\partial (\omega)^2}
{\it Re}\Sigma^R_{\dot{S}\dot{S}}({\bf k},\omega)|_{\omega=0}}
$$
Calculating $Z$ in fluctuation region results in the following expressions
for renormalization constant:
\begin{equation}
Z({\bf k} \to 0)=\frac{1}{1+\epsilon^{'}(k\xi)^2},\;\;\;\;
Z({\bf q} \to 0)=\frac{1}{1+\delta^{'}+\delta^{''}(q\xi)^2}
\label{3.3}
\end{equation}
where the coefficients $\epsilon^{'},\delta \ll 1$ also can be expressed in terms
of the integrals of different powers of static correlator $G_0$.
Using the definition (\ref{3.2}) for small, but nonzero frequencies one can obtain
the expansions of {\it real} spin diffusion coefficient $D^*$
and relaxation constant $\Gamma^*$
\footnote{$\omega^* \sim T_c\tau^{\nu z}$ is characteristic
fluctuations energy,
$z=3/2$}:
$$
D^{(2)*}({\bf k},\omega)=D_0(0,0)[1+\alpha^{'}(k\xi)^2 
+\alpha^{''}_{k\xi}
(\omega/\omega^*)^2 + ...]
$$
\begin{equation}
\Gamma^{(2)*}({\bf q},\omega)=\Gamma_0(0,0)[\beta+
\beta^{'}(q\xi)^2 +\beta^{''}_{k\xi}
(\omega/\omega^*)^2 + ...]
\label{3.5}
\end{equation}

It should be noted that we do not intend to  describe the kinetic
coefficients properties in the region $\omega \sim \omega^*$, 
$k,q \sim \xi^{-1}$. The  diffusion and relaxation picture in this
region becomes very complicated and scarcely amenable to
detailed analysis at the present time. For this reason we shall neglect
the irregular corrections to kinetic coefficients resulting from the 
generation of poles chain and drift of the poles toward the new cut 
\cite{Mal1},\cite{Mal2}. In the region  of ${\bf k}$, $\omega$ under consideration,
all these singularities are small corrections, and therefore, the phenomena
associated with them are no special interest.

Now we turn to  discussion of  the role of  many - particles 
 $(m >2)$ intermediate
states. As it was mentioned above, we are interested only in regular 
contributions of such diagrams to the kinetic coefficients:
$$
\frac{{\it Im}\Sigma^{R(m)}_{\dot{S}\dot{S}}}{\omega}\sim (ka)^2
\sum_{\bf p_1} ...\sum_{\bf p_m}
{\bf \Lambda^{(m)}}({\bf k},{\bf p}_1,...,{\bf p}_m )
{\bf \Lambda^{(m)}}^\dagger({\bf k},{\bf p}_1,...,{\bf p}_m )
\delta({\bf p_1 +...p_m -k})\times
$$

\begin{equation}
\times
\frac{1}{\pi^{m-1}}
\int_{-\infty}^{\infty}...\int_{-\infty}^{\infty}
\frac{d\varepsilon_1 ...d\varepsilon_m
{\it Im}K^R_{SS}({\bf p}_1,\varepsilon_1)...
{\it Im}K^R_{SS}({\bf p}_m,\varepsilon_m)}
{\varepsilon_1...\varepsilon_m}\delta(\varepsilon_1+...+
\varepsilon_m - \omega),
\label{3.4}
\end{equation}
Here the functions $K$ correspond both to "diffusons" and "relaxons". The
integration over frequency is still restricted  by the regions close to
singularities of scaling function. In the case of $m=2$ the
expression (\ref{3.4}) simply transforms into (\ref{2.7},\ref{2.8}).

The vertices ${\bf \Lambda^{(m)}}$ are symmetric functions of the momenta of
intermediate particles. For ${\bf k}=0$ the generalized Ward identities 
similar to thous of ref.
\cite{Mal1} hold for ${\bf \Lambda^{(m)}}$; therefore, 
${\bf \Lambda^{(m)}}$ can be expressed as a sum of derivatives 
of the ordinary n - particle vertices  of static scaling theory.
A "dimensional" estimate exists for the latter \cite{Mig68}:
$\Gamma_m \sim p^{3-m/2}$ in the limit ${\bf k} \to 0$. As a result, 
the diagrams with m-particle intermediate states in the channel of 
"diffusons" and "relaxons" creation do not change the scaling dimension
of irreducible self - energies. As to the vertices behavior
around the AF vector $Q$, there is a direct mapping of the properties
of the diagrams with two - particle intermediate states  onto those with many - 
 particle
intermediate states. Therefore, we can confine ourselves only 
by the diagrams with two - particle intermediate states. Another diagrams
only  renormalize the numerical constants, and the latter can not be
determined in a framework of our method. The energy dependence of vertices
also can not change the scaling behavior of kinetic coefficients 
 \cite{Mal1}.

\section*{Conclusion}

We investigated the scaling behavior of generalized kinetic
coefficients in 3D isotropic Heisenberg antiferromagnets. In a
framework of analysis based on the modified mode - mode coupling
theory the  critical exponents are derived. 
Necessary approximations in microscopic approach to satisfy the dynamic
scaling hypothesis are formulated. 
In particular, it is shown that the main sequence of
Feynman diagrams contain the processes with two - particles intermediate 
states. Besides, the vertices of static scaling theory are proved
to give the main contribution to quasiparticles scattering processes.

The momentum and frequency dependence of spin diffusion
and relaxation coefficients is found in the "pole" approximation.
A definition of the generalized effective coefficient is introduced
in a full analogy with that used in the theory of quantum liquids for a  definition of effective
mass. Taking into account the renormalization processes responsible for
multiple scattering of "diffusons" and "relaxons" we derived the 
asymptotics of scaling function in the region of small frequencies and 
momenta,  $\omega \ll \omega^*$,  $k,q \ll \xi^{-1}$.

Our investigation of the critical dynamic is based on the static and dynamic
scaling hypotheses. We took into account only two
coupling modes. One of them is considered to be responsible for
fluctuations of the total magnetization
and another deals with the fluctuations of order parameter - staggered
magnetization. Existence of two modes is determined by the conservation's laws
of the Heisenberg model. For this reason, our results  depend
only weakly on the specific properties of AF and we expect
that all they are  valid for any system
with nonconserved order parameter when the additional conserved quantity
exists.

The scaling behavior of more complicated systems, e.g. Ce - based
heavy fermion compounds with almost integer valence, can be significantly
different from critical dynamic of Heisenberg magnets. In these systems
an additional coupling modes interacting with 
paramagnons can appear due to, e.g., the  spin liquid correlations,
proposed in Ref.\cite{KKM94}. An additional modes in Kondo lattices
can be also connected with itinerant -  magnetic excitations. 
Kinetic coefficients of such systems
can be measured by the methods of  of inelastic neutron scattering
to  verify a spin liquid correlations existence.

The methods developed in this paper can be useful for  the systems
close to the quantum critical point \cite{Mil93},\cite{Hertz75} -- 
\cite{Mor88} when the temperatures of AF ordering are close to zero, or, even
negative. Our method also should be valid for anisotropy magnets, antiferrimagnets
and systems with dipolar interactions.

We think that the diagrammatic approach for investigation of
kinetic coefficient in the vicinity of Neel point has some advantages
in comparison with other methods \cite{Halp1}-\cite{Halp2},
\cite{Kaw76}. This approach can be applied to the  unrenormalizible
Hamiltonians and to systems with nonlocal coupling modes \cite{KKM97}.

\section*{Acknowledgments}

We would like to thank D.N.Aristov, Yu.Kagan, A.V.Lazuta, S.V.Maleev 
and V.L.Pokrovskii for valuable discussions. This work was supported by
Russian Foundation for Basic Research (95-02-04250a), International 
Association INTAS (93-2834) and Dutch Organization for Basic Research
NWO (07-30-002).

\begin{figure}
\begin{center}
\epsfig{%
file=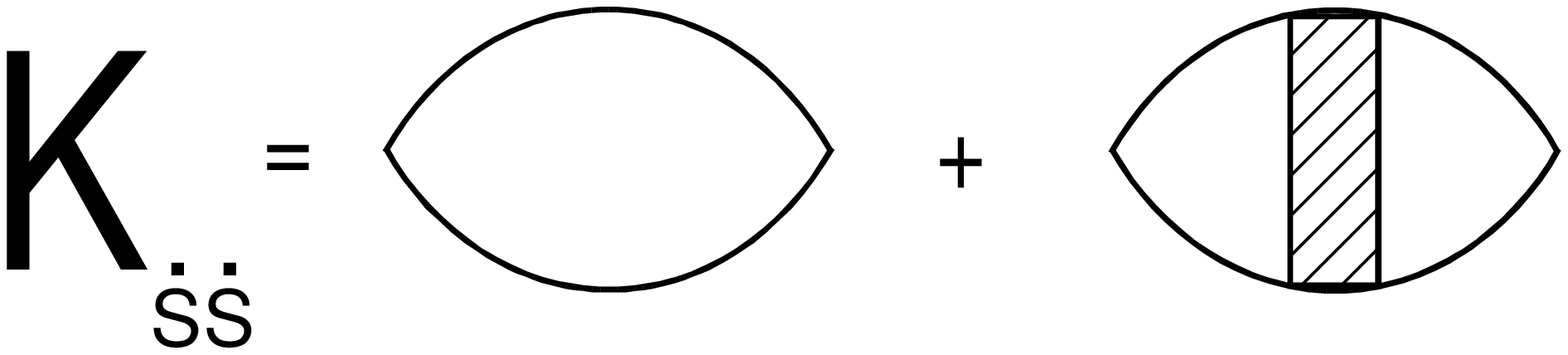,%
figure=fig1.eps,%
height=6cm,%
width=6cm,%
angle=0,%
}
\end{center}
\caption{Diagrammatic equation for spin - current correlator.}
\label{f1}
\end{figure}
\begin{figure}
\begin{center}
\epsfig{%
file=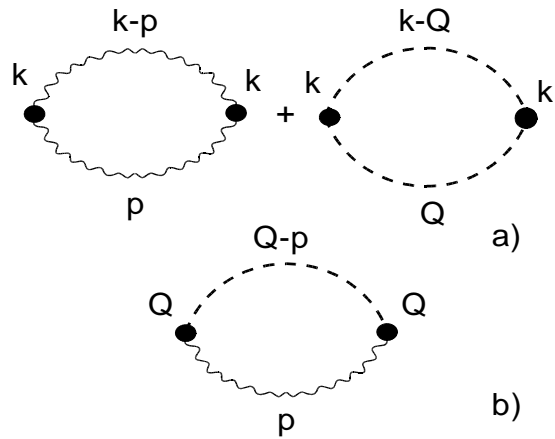,%
figure=fig2ab.eps,%
height=6cm,%
width=9cm,%
angle=0,%
}
\end{center}
\caption{Feynman diagrams for kinetic coefficients. Only two - particles intermediate 
states are taken into account. The diffusion mode is represented by the wavy
lines. Dashed lines stand for the relaxation mode. 
Dots represent the vertex parts from static scaling theory.}
\label{f2}
\end{figure}
\begin{figure}
\begin{center}
\epsfig{%
file=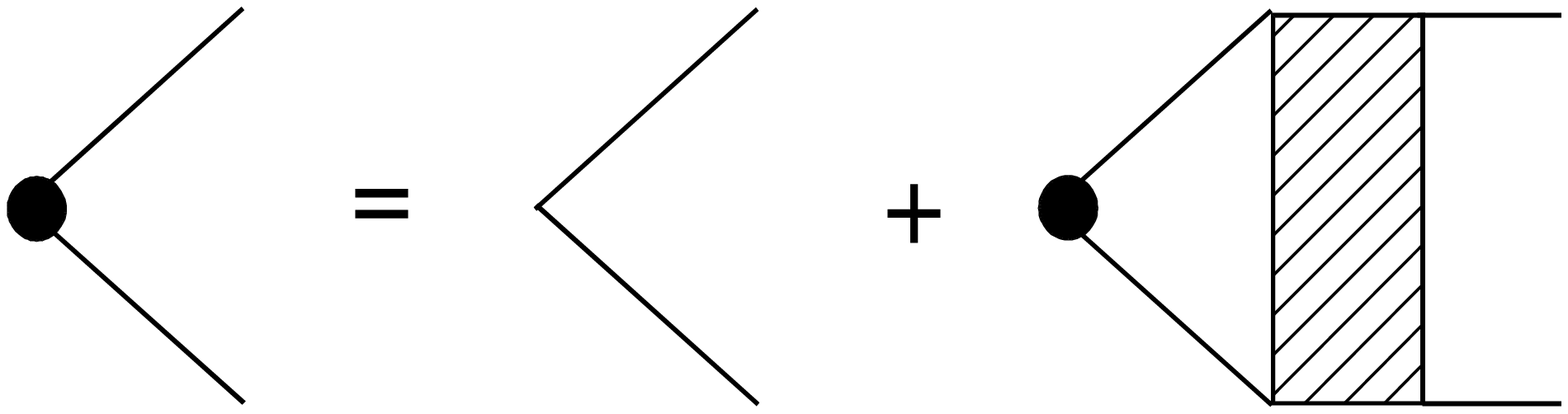,%
figure=fig3.eps,%
height=6cm,%
width=6cm,%
angle=0,%
}
\end{center}
\caption{Equations for two - particle dynamic vertex.}
\label{f3}
\end{figure}

\end{document}